\documentclass[10pt,conference,letterpaper]{IEEEtran}
\usepackage[left=0.625in,right=0.625in,top=0.75in,bottom=1in]{geometry}
\makeatletter
\def\ps@headings{%
\def\@oddhead{\mbox{}\scriptsize\rightmark \hfil \thepage}%
\def\@evenhead{\scriptsize\thepage \hfil \leftmark\mbox{}}%
\def\@oddfoot{}%
\def\@evenfoot{}}
\makeatother
\usepackage{graphicx,cite}
\usepackage[font=footnotesize]{caption}
\usepackage[font=footnotesize]{subcaption}

\usepackage{amssymb,amsmath,amsbsy}

\usepackage{algorithm,algorithmic}

\usepackage{array}

\usepackage{balance}

\usepackage[hyphens]{url}

\def\squareforqed{\hbox{\rlap{$\sqcap$}$\sqcup$}}
\def\qed{\ifmmode\squareforqed\else{\unskip\nobreak\hfil
\penalty50\hskip1em\null\nobreak\hfil\squareforqed
\parfillskip=0pt\finalhyphendemerits=0\endgraf}\fi}

\begin{document}

\title{RepFlow: Minimizing Flow Completion Times with Replicated Flows in Data Centers}
\author{
\IEEEauthorblockN{Hong Xu\IEEEauthorrefmark{1}\IEEEauthorrefmark{2}, Baochun Li\IEEEauthorrefmark{2}}
\IEEEauthorblockA{henry.xu@cityu.edu.hk, bli@eecg.toronto.edu}
\IEEEauthorblockA{ \IEEEauthorrefmark{1}
Department of Computer Science, City University of Hong Kong}
\IEEEauthorblockA{ \IEEEauthorrefmark{2}
Department of Electrical and Computer Engineering, University of Toronto}
}
\maketitle

\begin{abstract}

Short TCP flows that are critical for many interactive applications in data centers are plagued by large flows and head-of-line blocking in switches. Hash-based load balancing schemes such as ECMP aggravate the matter and result in long-tailed flow completion times (FCT). Previous work on reducing FCT usually requires custom switch hardware and/or protocol changes. We propose RepFlow, a simple yet practically effective approach that replicates each short flow to reduce the completion times, without any change to switches or host kernels. With ECMP the original and replicated flows traverse distinct paths with different congestion levels, thereby reducing the probability of having long queueing delay. We develop a simple analytical model to demonstrate the potential improvement of RepFlow. Extensive NS-3 simulations and Mininet implementation show that RepFlow provides 50\%--70\% speedup in both mean and 99-th percentile FCT for all loads, and offers near-optimal FCT when used with DCTCP.

\end{abstract}

\section{Introduction}
\label{sec:intro}

Data centers run many interactive services, such as search, social networking, and retail, that impose unique and stringent requirements on the transport fabrics. They often partition computation into many small tasks, distribute them to thousands of machines, and stitch the responses together to return the final result \cite{AGMP10,ZDMB12}. Such partition-aggregation workflows generate a large number of short query and response flows across many machines, and demand that short flows have low latency in order to provide soft real-time performance to users. More importantly, the tail latency also needs to be low since the request completion time depends on the slowest flow.

\begin{figure}[h]
\begin{minipage}[h]{0.48\linewidth}
	\centering
	\includegraphics[width=1\linewidth]{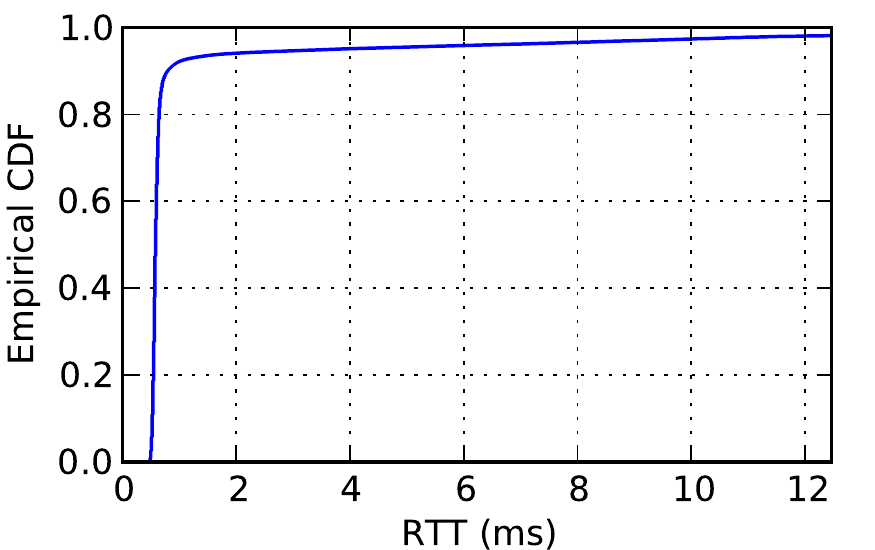}
	\vspace{-4.5mm}
	\caption{CDF of RTT between two small instances in EC2 us-west-2c.}
	\label{fig:meanrtt}
\end{minipage}
\begin{minipage}[h]{0.48\linewidth}
	\centering
	\includegraphics[width=1\linewidth]{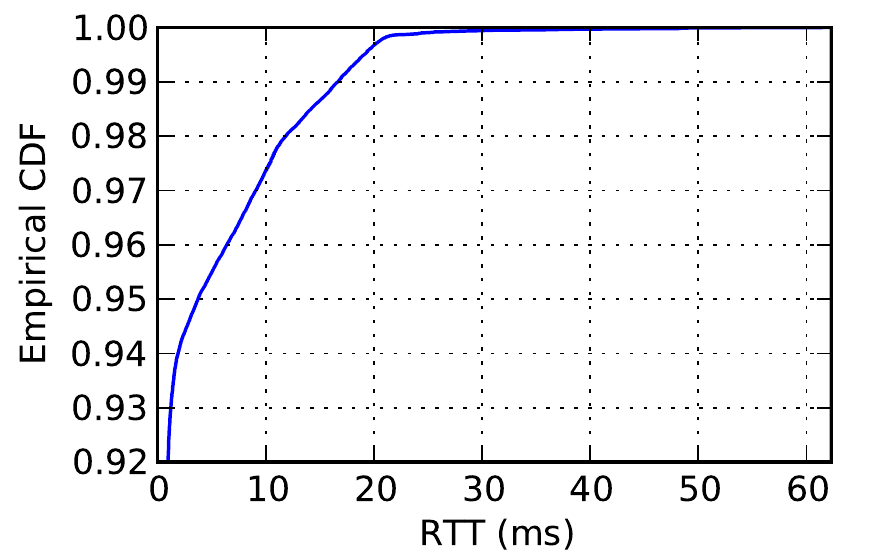}
	\vspace{-4.5mm}
	\caption{Long tail of the RTT distribution.}
	\label{fig:tailrtt}
\end{minipage}
\vspace{-1mm}
\end{figure}

TCP is the dominant transport protocol in data centers \cite{AGMP10}. Flow completion times (FCT) for short flows in TCP are poor: FCT can be as high as tens of milliseconds while in theory they could finish in 10--20 microseconds with 1G or 10G interconnects. The reason is that these flows often find themselves queued up behind bursts of packets from large flows of other workloads, such as backup, data mining, etc. The situation is even worse with imperfect load balancing schemes such as ECMP that perform hash-based flow-level load balancing among links of equal distance \cite{ARRH10}. ECMP is agnostic to congestion, does not differentiate between short and long flows, and may direct many large flows to the same path causing flash congestions and long-tailed FCT even when the network is lightly loaded \cite{ARRH10,ZDMB12}. We measure the round-trip times (RTT) between two small instances in Amazon EC2's us-west-2c zone every 1 second for 100K samples as a rough estimation of FCT. The newer EC2 data centers are known to have many equal-cost paths between given pairs of instances \cite{RBPG11}. Fig.~\ref{fig:meanrtt} and Fig.~\ref{fig:tailrtt} confirm the long-tailed distribution: While mean RTT is only 0.5ms, the 99-th percentile RTT is 17ms.

Recent research has proposed many new transport designs to reduce FCT of short flows. Broadly speaking, the central idea is to reduce the short flows' queueing delay via adaptive ECN-based congestion control \cite{AGMP10}, explicit flow rate assignment \cite{HCG12}, deadline-aware scheduling \cite{WBKR11,VHV12,MQUM13}, priority queueing \cite{AYSK13}, and cross-layer redesign \cite{ZDMB12,AYSK13}. While effective, they do require modifications to switches and operating systems, which makes them difficult to deploy in a large-scale data center with a large number of servers and switches.

Our goal in this paper is to design a practical and effective data center transport scheme that provides low latency for short flows both on average and in the 99-th percentile, and can be readily deployed in current infrastructures. To this end, we present RepFlow, a simple data center transport design. RepFlow directly uses existing TCP protocols deployed in the network. The only difference with RepFlow is that it replicates each short TCP flow by creating another TCP connection to the receiver, and sending identical packets for both flows. The application uses the first flow that finishes the transfer. Flow replication can be easily implemented as libraries or middleware at the application layer. Thus RepFlow requires no change to switches, the network stack, or operating systems, and maintains TCP's robustness for throughput intensive traffic. It can also be used with other data center transport protocols such as DCTCP \cite{AGMP10} to further improve performance as we will show later.

The key conceptual insight behind RepFlow is the observation that multi-path diversity, which is readily available with high bisection bandwidth topologies such as Fat-tree \cite{ALV08}, is an effective means to combat performance degradation that happens in a random fashion. Flash congestion and queueing delay due to bursty traffic and ECMP happen randomly in any part of the network at any time. As a result, congestion levels along different paths are statistically independent. In RepFlow, the replicated and original flow are highly likely to traverse different paths, and the probability that both experience long queueing delay is much smaller. 

It is important to note that RepFlow is built atop ECMP with per-flow load balancing, and is different from multipathing schemes such as MPTCP \cite{RBPG11} and packet spraying \cite{DPHK13} that split a flow across multiple paths. Traffic is asymmetric and dynamic, especially considering link failures and external traffic that origintes or terminates outside of the data center. When the paths used by a flow have different loads, out-of-order packets interact negatively with TCP and splitting flows hardly achieves latency gains for short flows, though it may improve the aggregate throughput for large flows. ECMP is also widely used in current data centers, reducing the implementation overhead of RepFlow.



We evaluate our design of RepFlow with queueing analysis, detailed packet-level simulations in NS-3, and Linux kernel-based implementation using Mininet \cite{HHJL12}. We develop a simple M/G/1 queueing model to model mean and tail FCT. Our model shows that the diversity gain of replication can be understood as a reduction in the effective traffic load seen by short flows, which leads to significantly improved queueing delay and FCT. Our evaluation uses two widely used data center workloads: one that mimics a web search workload \cite{AGMP10} and one that mimics a typical data mining workload \cite{GHJK09}. NS-3 simulations with a 16-pod 1,024-host Fat-tree, and experiments with a 4-pod Fat-tree on Mininet both show that RepFlow achieves 50\%--70\% speedup in both mean and 99-th percentile FCT even for loads as high as 0.8 compared to TCP. When it is feasible to use advanced transport protocols such as DCTCP \cite{AGMP10}, RepFlow offers competitive performance compared to state-of-the-art clean slate approaches such as pFabric \cite{AYSK13}. The overhead to the network is negligible, and large flows are virtually not impacted. Thus we believe it is a lightweight and effective approach that requires minimal implementation efforts on top of existing TCP based transport layer with salient FCT reductions. 


\section{Related Work}
\label{sec:related}

Motivated by the drawbacks of TCP, many new data center transport designs have been proposed. We briefly review the most relevant prior work in this section. We also introduce some additional work that uses replication in wide-area Internet, MapReduce, and distributed storage systems for latency gains. 

{\bf Data center transport.} DCTCP \cite{AGMP10} and HULL \cite{AKEP12} use ECN-based adaptive congestion control and appropriate throttling of large flows to keep the switch queue occupancy low in order to reduce short flows' FCT. D$^3$ \cite{WBKR11}, D2TCP \cite{VHV12}, and PDQ \cite{HCG12} use explicit deadline information to drive the rate allocation, congestion control, and preemptive scheduling decisions. DeTail \cite{ZDMB12} and pFabric \cite{AYSK13} present clean-slate designs of the entire network fabric that prioritize latency sensitive short flows to reduce the tail FCT. All of these proposals require modifications to switches and operating systems. Our design objective is different: we strive for a simple way to reduce FCT without any change to TCP and switches, and can be readily implemented at layers above the transport layer. RepFlow presents such a design with simple flow replication that works with any existing transport protocol.

{\bf Replication for latency.} Though seemingly naive, the idea of using replication to improve latency has in fact gained increasing attention in both academia and industry as a general technique for its simplicity and effectiveness. Google reportedly uses request replications to rein in the tail response times in their distributed systems \cite{D12}. In the context of wide-area Internet, \cite{VMGS12} argues for the use of redundant operations to improve latency of DNS queries, and \cite{WM13} argues for the latency benefit of having multiple wide-area transit links in a multi-cloud CDN deployment scenario. Replication has also been recently used in MapReduce \cite{AGSS13} and storage systems \cite{SCG13} to mitigate straggling jobs.

\section{RepFlow: Motivation and Design}
\label{sec:design}

\subsection{Motivation}

RepFlow's key design insight is that multi-path diversity exists profoundly as a result of randomized load balancing. In today's data center networks based on a Fat-tree or Clos topology \cite{AGMP10,GHJK09}, many paths of equal distance exist between a given pair of end-hosts. Equal-cost multi-path routing, or ECMP, is used to perform flow-level load balancing. When a packet arrives at a switch, ECMP picks an egress port uniformly at random among equal-cost paths based on the hash value of its five-tuple in the packet header. All packets of the same flow then follow a consistent path. 
\begin{figure}[h]
	\centering
	\includegraphics[width=0.7\linewidth]{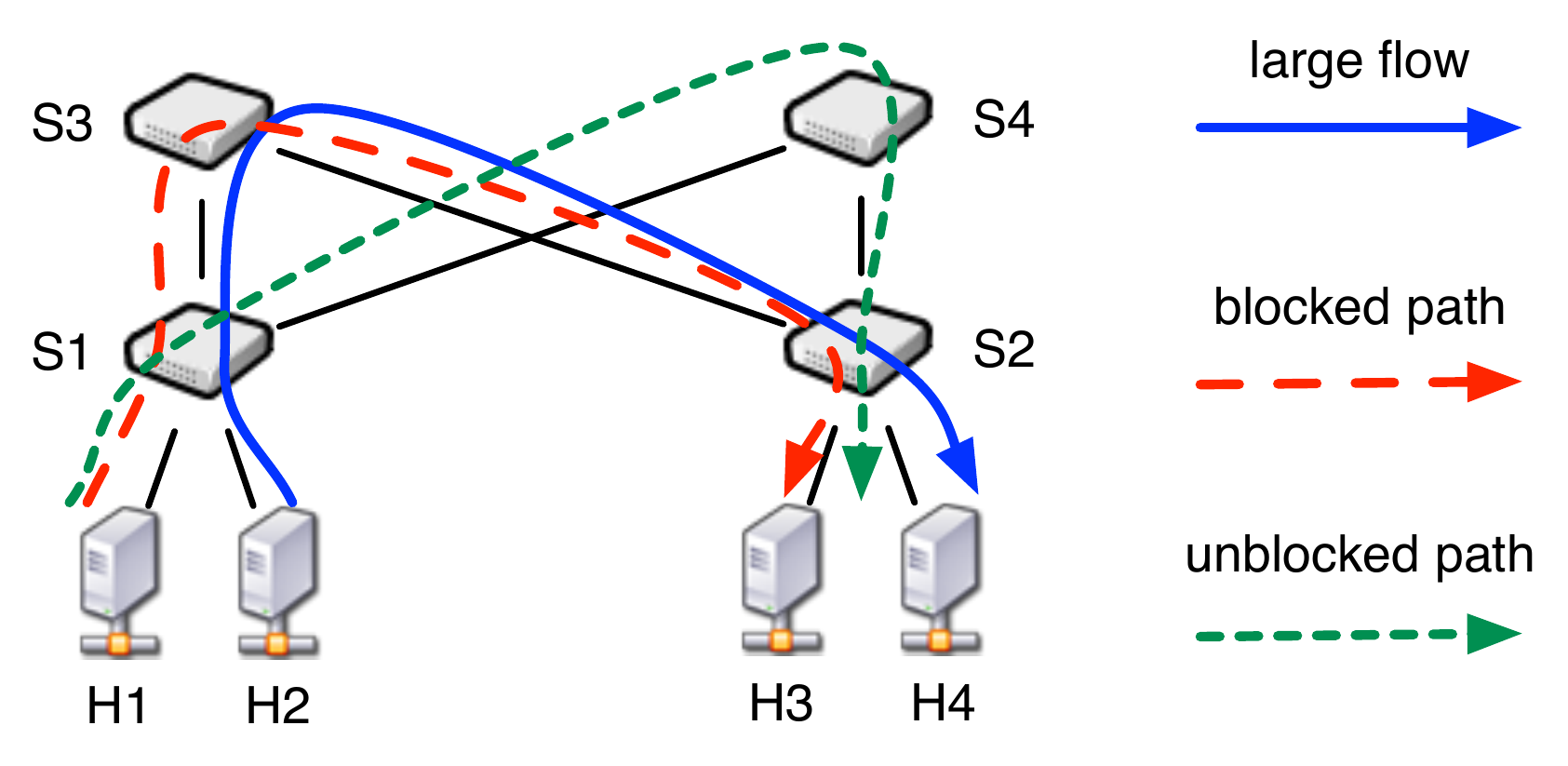}
	\vspace{-1.5mm}
	\caption{A large flow transmits from H2 to H4 following the path shown in the solid line. If the short flow from H1 to H3 takes the path S1--S3--S2, it will queue behind packets of the large flow in the links S1--S3 and S3--S2. If it takes the path S1--S4--S2, there is no head-of-line blocking since all ports of this path are empty. ECMP will randomly hash a short flow to one of the two paths.}
	\label{fig:toytopo}
\end{figure}

Due to its randomness, it is often the case that short and large flows are routed on the same path. Short flows then have to wait until packets of large flows are processed, a phenomenon known as head-of-line blocking. Consider a toy example shown in Fig.~\ref{fig:toytopo}. The topology resembles one pod of a 4-pod Fat-tree network. A host under switch S1(S2) has two paths to each host under switch S2(S1). There is a persistent large flow from H2 to H4, taking the path S1--S3--S2. Now H1 starts to send short flows of 10 packets continuously to H3. ECMP will randomly hash the short flows with 0.5 probability to the path S1--S3--S2, which results in head-of-line blocking and long FCT. We conduct an experiment in Mininet \cite{HHJL12} using exactly the same setup, and observe the mean FCT is 10x worse when the large flow co-exists as shown in the following table (more on Mininet in Sec.~\ref{sec:mininet}). 
\begin{table}[htp]
    \centering
    \begin{tabular}{ c | c | c }
    \hline 
	    Scenario & Mean FCT & 99-th percentile FCT  \\ \hline
	    without the large flow & 0.0135s & 0.0145s \\ \hline
	    with the large flow & 0.175s & 0.490s \\ \hline
	    with the large flow and replication & 0.105s & 0.212s \\ 
    \hline  
    \end{tabular}
    \caption{Mininet experiment results of the toy example shown in Fig.~\ref{fig:toytopo}. Each link is 50Mb with 1ms delay. Short flows of 10 packets are sent continuously from H1 to H3 with ECMP. With the large flow, short flows suffer from 10x worse FCT. Replication dramatically improves mean and tail FCT by over 50\% in this case.}
\label{table:toy}
\vspace{-2mm}
\end{table}

Thus ECMP creates serious problems. Yet at the same time it also provides a promising solution to the problem --- multi-path diversity, which motivates RepFlow. In the toy example it is obvious that the path S1--S4--S2 has much less queueing delay, and by replicating the short flow and making sure the two flows have distinct five-tuples, one copy of it will traverse this path and improve FCT significantly. From the same Mininet experiment we do observe over 50\% improvement with simple replication in this case, as shown in Table~\ref{table:toy}. Note in reality it is in general difficult to know and choose the right path for short flows beforehand in a data center network with a number of flows, not to mention its latency overhead. Replication removes this need by opportunistically utilizing the less congested paths. 

\subsection{Design}

RepFlow uses flow replication to exploit multi-path diversity. It does not modify the transport protocol, and thus works on top of TCP, and evidently any other TCP variants, such as DCTCP \cite{AGMP10} and D2TCP \cite{VHV12}. On the high level, there are several design choices we need to make. First, which short flow should be replicated? We heuristically mandate that flows less than or equal to 100KB are considered short flows, and are replicated to achieve better latency. This threshold value is chosen in accordance with many existing papers \cite{ZDMB12,HCG12,AYSK13,MQUM13}. Second, we need to decide when to replicate the flows. One might argue that we should only replicate the flows when they are experiencing long queueing delays so the replication overhead is reduced. However the short duration of these flows makes such a reactive approach too slow to possibly remedy the situation. In the current design RepFlow proactively replicates each and every short flow from the very beginning to achieve the best latency. As we will show in Sec.~\ref{sec:overhead}, the overhead of doing so is negligible, thanks to the well-known fact that short flows only account for a tiny fraction of total bytes in production systems \cite{AYSK13,KSGP09}. Finally, we replicate exactly once for simplicity, though more replication is possible.

RepFlow can be implemented in many ways. The simplest is to create two TCP sockets when a short flow arrives, and send the same packets through two sockets. This is also our current implementation. Since data centers run a large number of applications, it is preferable to provide RepFlow as a general library or middleware for any application to invoke \cite{pp}. For example one may implement RepFlow as a new transport abstraction in Thrift, a popular RPC framework used by companies like Facebook \cite{SAK07}. We are currently investigating this option. Another possibility is to implement RepFlow at the transport layer, by modifying TCP protocol and header so short flows are marked and automatically replicated with two independent subflows. This approach provides transparency to applications, at the cost of requiring kernel upgrades. In this space, RepFlow can be incorporated into MPTCP \cite{RPBF12} with its multi-path support.

It is evident that RepFlow lends itself to many implementation choices. No matter the details, it is crucial to ensure path diversity is utilized, i.e. the five-tuples of the original and replicated flow have to be different (assuming ECMP is used). In our implementation we use different destination port numbers for this purpose.

\section{Analysis}
\label{sec:model}

Before we evaluate RepFlow at work using simulations and experiments, in this section we present a queueing analysis of flow completion times in data centers to theoretically understand the benefits and overhead of replication.

\subsection{Queueing Model}
\label{sec:queueing_model}

A rich literature exsits on TCP steady-state throughput models for both long-lived flows \cite{PFTK98,MSMO97} and short flows \cite{HOT97}. There are also efforts in characterizing the completion times of TCP flows \cite{CSA00, LBS11}. See \cite{CSA00} and references therein for a more complete literature review. These models are developed for wide-area TCP flows, where RTTs and loss probabilities are assumed to be constants. Essentially, these are open-loop models. The data center environment, with extremely low fabric latency, is distinct from the wide-area Internet. RTTs are largely due to switch queueing delay caused by TCP packets, the sending rate of which in turn are controlled by TCP congestion control reacting to RTTs and packet losses. This closed-loop nature makes the analysis more intriguing \cite{PD08}. 

Our objective is to develop a simple FCT model for TCP flows that accounts for the impact of queueing delay due to large flows, and demonstrates the potential of RepFlow in data center networks. We do not attempt to build a fine-grained model that accurately predicts the mean and tail FCT, which is left as future work. Such a task is potentially challenging because of not only the reasons above, but also the complications of timeouts and retransmissions \cite{PKVA08,VPSK09}, switch buffer sizes \cite{AKM04,LBS11}, etc. in data centers. 

We construct our model based on some simplifying assumptions. We abstract one path of a data center network as a M/G/1 first-come-first-serve (FCFS) queue with infinite buffer. Thus we do not consider timeouts and retransmissions. Flows arrive following a Poisson process and have size $X\sim F(\cdot)$. Since TCP uses various window sizes to control the number of in-flight packets, we can think of a flow as a stream of bursts arriving to the network. We assume the arrival process of the bursts is also Poisson. One might argue that the arrivals are not Poisson as a burst is followed by another burst one RTT later (implying that interarrival times are not even i.i.d). However queueing models with general interarrival time distributions are difficult to analyze and fewer results are available \cite{F11a}. For tractability, we rely on the commonly accepted M/G/1-FCFS model \cite{AKM04,AYSK13}. We summarize some key notations in the table below. Throughout this paper we consider (normalized) FCT defined as the flow's completion time normalized by its best possible completion time without contention. 

\begin{table}[htp]
    \centering
    \caption{Key notations.}
    \begin{tabular}{ c | c  }
    \hline 
	    $M$ & maximum window size (64KB, 44 packets) \\ \hline
	    $S_L$ & threshold for large flows (100KB, 68 packets) \\ \hline
	    $F(\cdot), f(\cdot)$ & flow size CDF and PDF \\ \hline
        $\rho\in[0,1)$ & overall traffic load \\ \hline
        $W$ & queueing delay of the M/G/1-FCFS queue \\ \hline
        $k$ & initial window size in slow-start \\ \hline
    \end{tabular}
\label{table:notations}
\end{table}

For short flows, they mostly stay in the slow-start phase for their life time \cite{CSA00,AKM04,DRCC10,LBS11}. Their burst sizes depend on the initial window size $k$. In slow-start, each flow first sends out $k$ packets, then $2k$, $4k$, $8k$, etc. Thus, a short flow with $X$ packets will be completed in $\log_2 (X/k+1)$ RTTs, and its normalized completion time can be expressed as
\begin{equation}\label{eqn:fct_short}
	FCT_X = \sum_{i=1}^{\log_2 (X/k+1)} W_i/X + 1,
\end{equation}
assuming link capacity is 1 packet per second.

For large flows that are larger than $S_L$, we assume that they enter the congestion avoidance phase immediately after it arrives \cite{PFTK98,MSMO97}. They continuously send bursts of a fixed size equal to the maximum window size $M$ (64KB by default in Linux). A large flow's FCT is then
\begin{equation}\label{eqn:fct_large}
	FCT^L_X = \sum_{i=1}^{X/M} W_i/X + 1, X\ge S_L.
\end{equation}

\subsection{Mean FCT Analysis}
\label{sec:mean_FCT}

We now analyze the mean FCT for short flows in TCP. On average, each burst sees the expected queueing delay $E(W)$. Thus according to \eqref{eqn:fct_short}, the mean FCT of a flow with $X$ packets is
\begin{equation*}
	E[FCT_X] = \log_2 (X/k+1) \frac{E[W]}{X} + 1.
\end{equation*}
The mean FCT for short flows less than $S_L$ is
\begin{equation}\label{eqn:fct_mean}
	E[FCT] = E[W] \int_0^{S_L}\frac{\log_2 (x/k+1) }{x } \frac{f(x)}{F(S_L)} \mathrm{d}x + 1.
\end{equation}
The mean queueing delay of a M/G/1-FCFS queue with load $\rho$ is obtained with the famous Pollaczek-Khintchine formula \cite{GSTH08}:
\begin{equation}\label{eqn:pk}
	E[W] = \frac{\rho}{2(1-\rho)} \frac{E[B^2]}{E[B]} = \frac{\rho M}{2(1-\rho)},
\end{equation}
where $B$ denotes the burst size as opposed to the flow size. Since most of the bytes are from large flows, almost all bursts arrive to the queue are of a fixed size $M$, and $E[B^2]/E[B] = M$. Therefore we have
\begin{equation}
	E[FCT] = \frac{\rho M}{2(1-\rho)} \int_0^{S_L}\frac{\log_2 (x/k+1) }{x } \frac{f(x)}{F(S_L)} \mathrm{d}x + 1.
\end{equation}

\begin{figure}[hpt]
\begin{minipage}[h]{0.49\linewidth}
	\centering
	\includegraphics[width=1\linewidth]{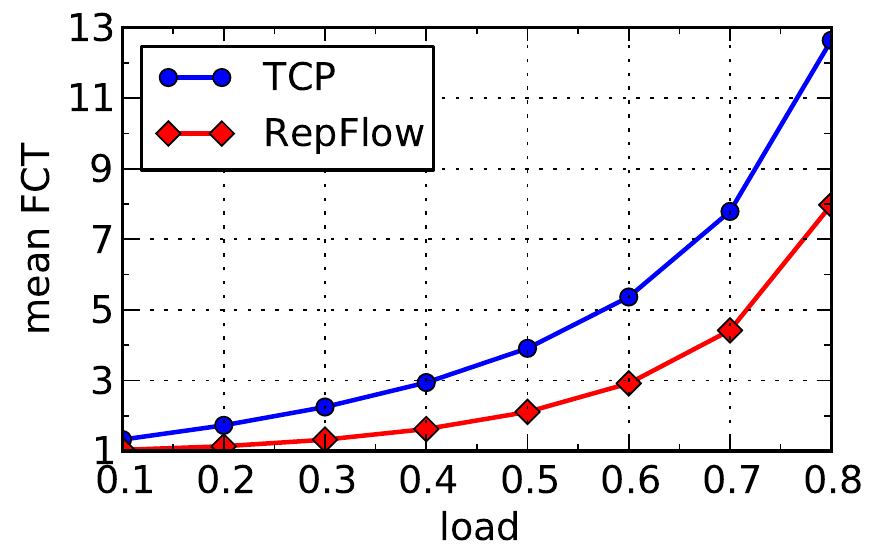}
	\vspace{-6.5mm}
	\caption{Short flow mean FCT. $k=12$ packets, flow size distribution from the web search workload \cite{AGMP10}.}
	\label{fig:meanFCT}
\end{minipage}
\begin{minipage}[h]{0.49\linewidth}
	\centering
	\includegraphics[width=1\linewidth]{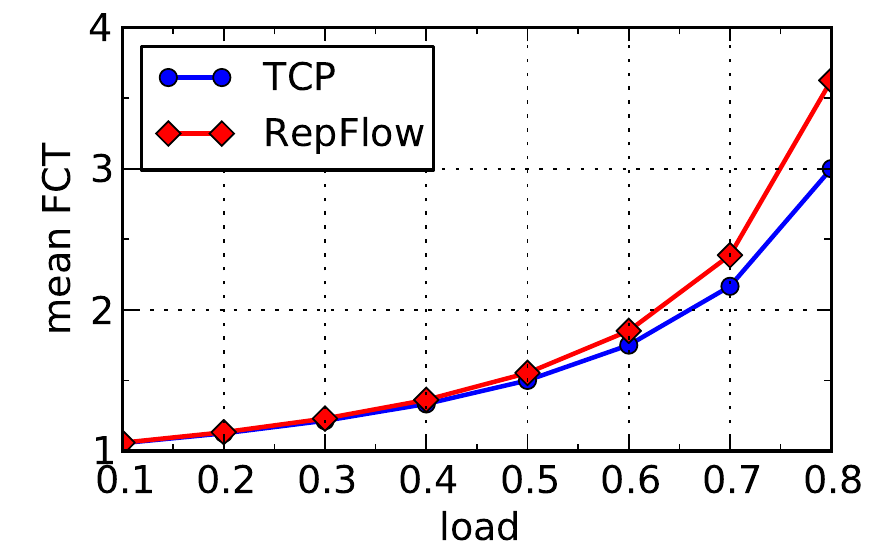}
	\vspace{-6.5mm}
	\caption{Large flow mean FCT. $k=12$ packets, flow size distribution from the web search workload \cite{AGMP10}.}
	\label{fig:meanFCT_large}
\end{minipage}
\vspace{-3mm}
\end{figure}

The mean FCT for short flows depends on the load of the network and the flow size distribution. Many data center operators opt for an increased initial window size to reduce latency in the slow-start phase \cite{DRCC10}. So we use $k=12$ packets \cite{DRCC10,AYSK13} throughout the paper. Using a flow size distribution from a production data center running web search workloads \cite{AGMP10}, Fig.~\ref{fig:meanFCT} plots the FCT with varying load. 

We now turn our attention to RepFlow, and obtain its mean FCT expression. For each short flow, RepFlow sends two identical copies by initiating two TCP connections between the same end-points. With ECMP, each flow is transmitted along different paths and experiences different congestion levels. We model this as having two independent queues with independent arrival processes and the same load $\rho$. When a short flow arrives, it enters both queues and get serviced, and its completion time is based on the faster queue. 

Without replication, each short flow sees a queue of load $\rho$, i.e. the network is busy with probablility $\rho$ when the flow enters, and idle with probablility $1-\rho$. Now with replication, each queue's load is slightly increased from $\rho$ to $(1+\epsilon)\rho$, where
\begin{equation*}
	\epsilon = \frac{\int_0^{S_L} x f(x)\mathrm{d}x }{E[X]}.
\end{equation*}
$\epsilon$ is the fraction of total bytes from short flows, and is usually very small (less than 0.1 \cite{AGMP10,GHJK09,KSGP09}). Since the two queues are independent, a short flow will find the network busy only when both queues are busy with probability $(1+\epsilon)^2\rho^2$, and idle with probability $1-(1+\epsilon)^2\rho^2$. In other words, each flow is effectively serviced by a virtual queue of load $(1+\epsilon)^2\rho^2$. Thus, the mean FCT for RepFlow is simply
\begin{align}
	E[FCT_{rep}]  = & \frac{(1+\epsilon)^2\rho^2 M}{2(1-(1+\epsilon)^2\rho^2)} \int_0^{S_L}\frac{\log_2 (x/k+1) }{x } \frac{f(x)}{F(S_L)} \mathrm{d}x \nonumber\\
	& + 1. \label{eqn:meanFCT_rep}
\end{align}
For small $\epsilon\le 0.1$, $(1+\epsilon)^2\rho^2$ is much smaller than $\rho$. As $\rho$ increases the difference is smaller. However the factor $\rho/(1-\rho)$ that largely determines the queueing delay $E[W]$ and FCT is very sensitive to $\rho$ in high loads, and a small decrease of load leads to significant decrease in FCT. In the same Fig.~\ref{fig:meanFCT}, we plot FCT for RepFlow with the same web search workload \cite{AGMP10}, where 95\% of bytes are from large flows, i.e. $\epsilon=0.05$. Observe that RepFlow is able to reduce mean FCT by a substantial margin compared to TCP in all loads.

Our analysis reveals that intuitively, the benefit of RepFlow is due to a significant decrease of effective load experienced by the short flows. Such a load reduction can be understood as a form of multi-path diversity discussed earlier as a result of multi-path network topologies and randomized load balacing.

At this point one may be interested in understanding the drawback of RepFlow, especially the effect of increased load on large flows. We now perform a similar FCT analysis for large flows. For a large flow with $X>S_L$ packets, substitute \eqref{eqn:pk} to \eqref{eqn:fct_large} yields
\begin{equation}\label{eqn:FCT_large}
	E[FCT^L] = \frac{\rho M}{2(1-\rho)} \frac{X}{M\cdot X} + 1 = \frac{\rho}{2(1-\rho)} + 1.
\end{equation}
The mean FCT for large flows only depends on the traffic load. With RepFlow, load increases to $(1+\epsilon)\rho$, and FCT becomes
\begin{equation}\label{eqn:FCT_rep_large}
	E[FCT^L_{rep}] = \frac{(1+\epsilon)\rho}{2(1-(1+\epsilon)\rho)} + 1,
\end{equation}
For large flows, load only increases by $\epsilon$, whereas small flows see a load decrease of $1-(1+\epsilon)^2\rho$. Large flows are only mildly affected by the overhead of replication. Fig.~\ref{fig:meanFCT_large} plots the mean FCT comparison for large flows. As we shall see from simulations and implementations results in Sec.~\ref{sec:overhead} and Sec.~\ref{sec:mininet} the performance degradation is almost negligible even in high loads.

\subsection{99-th Percentile FCT Analysis}
\label{sec:tailFCT}

To determine the latency performance at the extreme cases, such as the 99-th percentile FCT \cite{AGMP10,ZDMB12,AKEP12,HCG12}, we need the probability distribution of the queueing delay, not just its average. This is more difficult as no closed form result exists for a general M/G/1 queueing delay distribution. Instead, we approximate its tail using the effective bandwidth model \cite{K96}, which gives us the following
\begin{equation}\label{eqn:ld}
	P(W>w) \approx e^{ -w \frac{2(1-\rho)}{\rho}\cdot \frac{E[X]}{E[X^2]} } = e^{ -w \frac{2(1-\rho)}{\rho M} }.
\end{equation}
This equation is derived in the extended version of \cite{AKM04}. Setting \eqref{eqn:ld} equal to 0.01, we obtain the 99-th percentile queueing delay $\tilde{W}$:
\begin{equation}\label{eqn:tailwaiting}
	\tilde{W} = \ln 10\cdot \frac{\rho M}{1-\rho} = 2\ln 10\cdot E[W].
\end{equation}

\begin{figure}
\begin{minipage}[h]{0.49\linewidth}
	\centering
	\includegraphics[width=1\linewidth]{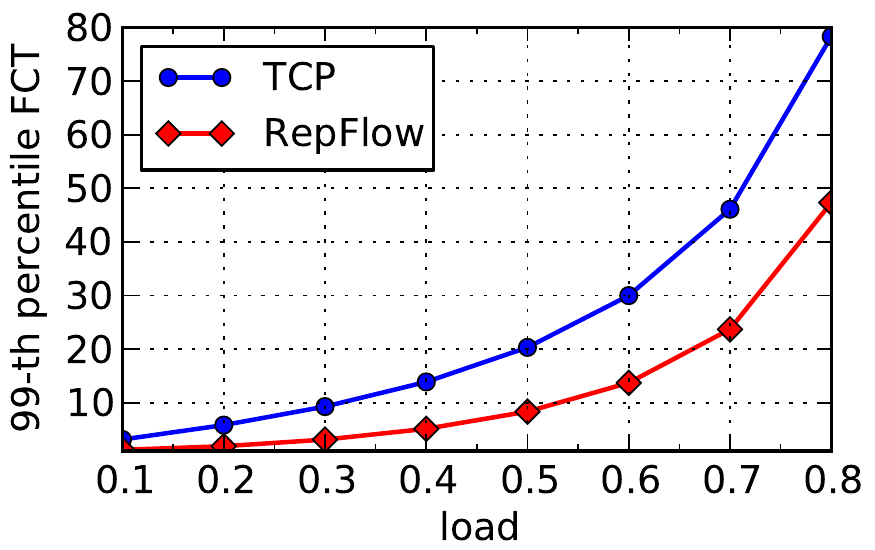}
	\vspace{-6.5mm}
	\caption{Short flow tail FCT. $k=12$ packets, flow size distribution from the web search workload \cite{AGMP10}.}
	\label{fig:tailFCT}
\end{minipage}
\vspace{-2mm}
\begin{minipage}[h]{0.49\linewidth}
	\centering
	\includegraphics[width=1\linewidth]{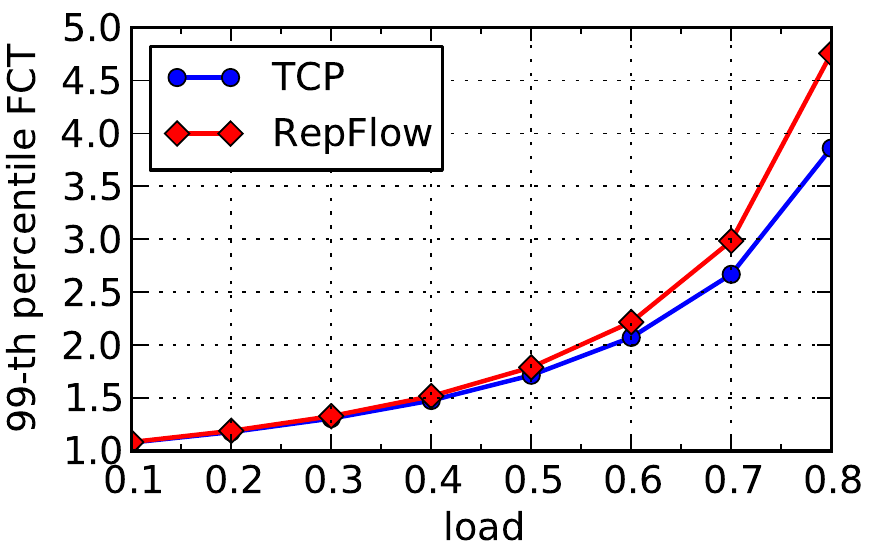}
	\vspace{-6.5mm}
	\caption{Large flow tail FCT. $k=12$ packets, flow size distribution from the web search workload \cite{AGMP10}.}
	\label{fig:tailFCT-large}
\end{minipage}
\vspace{-2mm}
\end{figure}

Recall short flows finish in $\log_2 (X/k+1) $ rounds. If a flow experiences a queueing delay of $\tilde{W}$ in one round, the total FCT will be guaranteed to hit the 99-th percentile tail\footnote{We cannot say that the delay is $\tilde{W}$ for all rounds, which happens with probability $0.01^{\log_2 (X/k+1)}$, i.e. much smaller than 0.01 even for small number of rounds.}. Thus, we can approximate the 99-th percentile FCT for short flows using $\tilde{W}$ as:
\begin{align}
	\tilde{FCT} & = E[W]\int_0^{S_L}\frac{\log_2 (x/k+1)-1+2\ln 10 }{x } \frac{f(x)}{F(S_L)} \mathrm{d}x + 1 \nonumber \\
	& = E[W]\cdot N + 1. \label{eqn:tailfct}
\end{align}
By the same token, we can calculate the 99-th percentile FCT for short flows under RepFlow.
\begin{align}
	\tilde{FCT_{rep}} & = E[W_{rep}]\cdot N + 1, \label{eqn:tailfct_rep} \\
	\text{where } & E[W_{rep}]  = \frac{(1+\epsilon)^2\rho^2 M}{2(1-(1+\epsilon)^2\rho^2)}. \nonumber
\end{align}
From \eqref{eqn:tailfct} and \eqref{eqn:tailfct_rep} we can see that the tail FCT depends critically on the queueing delay, which is determined by the traffic load $\rho$. Therefore RepFlow provides better tail latency in addition to better average latency, since it reduces the effective load seen by the short flows. Fig.~\ref{fig:tailFCT} shows the numerical results using the web search workload where we observe $\sim$40\%--70\% tail FCT improvement. 

According to queueing theory, the most likely reason for the extreme events such as 99-th percentile FCT to happen is that for some time all inter-arrival times are statistically smaller than usual \cite{BZ07}. This has an intuitive interpretation in our problem. Recall that our queue resembles a path of the data center network connecting many pairs of end-hosts. The arrival process to our queue is in fact a composition of many Poisson arrival processes generated by the hosts, with ECMP controlling the arrival rates. While the aggregate arrival rate on average is $\lambda$, at times the queue would see the instantaneous arrival rates from individual hosts much higher than usual due to hash collisions in ECMP, resulting in the tail FCT. 



The tail FCT analysis for large flows can be similarly derived as follows. 
\begin{align}
	\tilde{FCT^L} &= E[FCT^L] + (2\ln 10 -1) E[W]\cdot P,\label{eqn:tailfct-large} \\
	\tilde{FCT^L_{rep}} &= E[FCT^L_{rep}] + (2\ln 10 -1) E[W_{rep}^L]\cdot P,\label{eqn:tailfct-large-rep} \\
	& \text{where } P=\int^{\infty}_{S_L} \frac{1}{x} \frac{f(x)}{1-F(S_L)}\mathrm{d} x, \nonumber \\
	& E[W_{rep}^L] = \frac{(1+\epsilon)\rho M}{2(1-(1+\epsilon)\rho)}. \nonumber
\end{align}
Fig.~\ref{fig:tailFCT-large} shows the numerical results. Large flows enjoy better tail FCT performance compared to short flows, since their transmission lasts for a long time and is not sensitive to long-tailed queueing delay. Again observe that RepFlow does not penalize large flows.

\begin{figure*}[t]
\begin{subfigure}[t]{0.33\linewidth}
	\centering
	\includegraphics[width=1\linewidth]{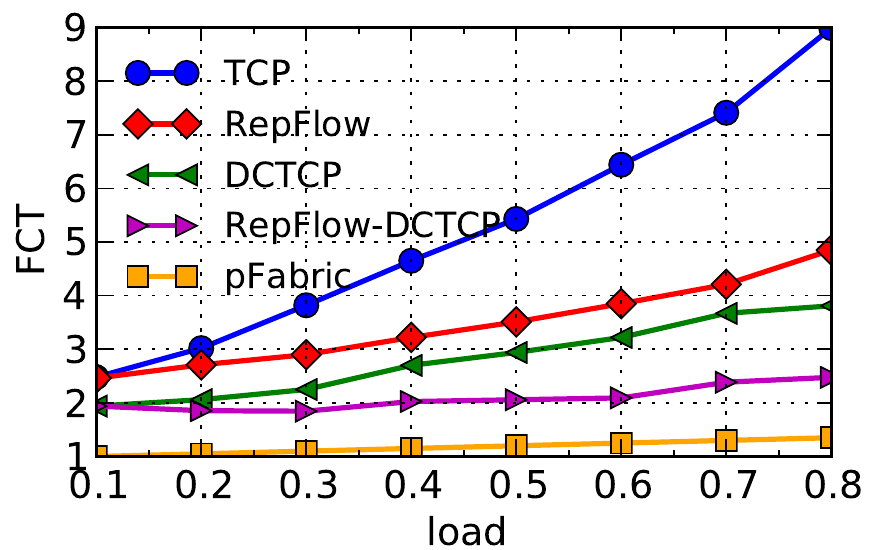}
	\vspace{-4.5mm}
	\caption{(0,100KB]: Avg}
	\label{fig:mean-short-web-sim}
\end{subfigure}
\begin{subfigure}[t]{0.33\linewidth}
	\centering
	\includegraphics[width=1\linewidth]{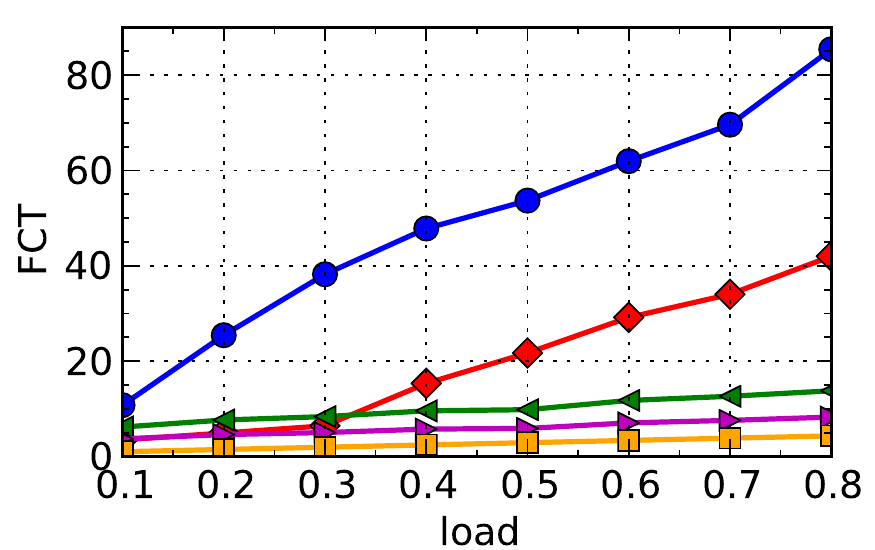}
	\vspace{-4.5mm}
	\caption{(0,100KB]: 99-th percentile}
	\label{fig:tail-short-web-sim}
\end{subfigure}
\begin{subfigure}[t]{0.33\linewidth}
	\centering
	\includegraphics[width=1\linewidth]{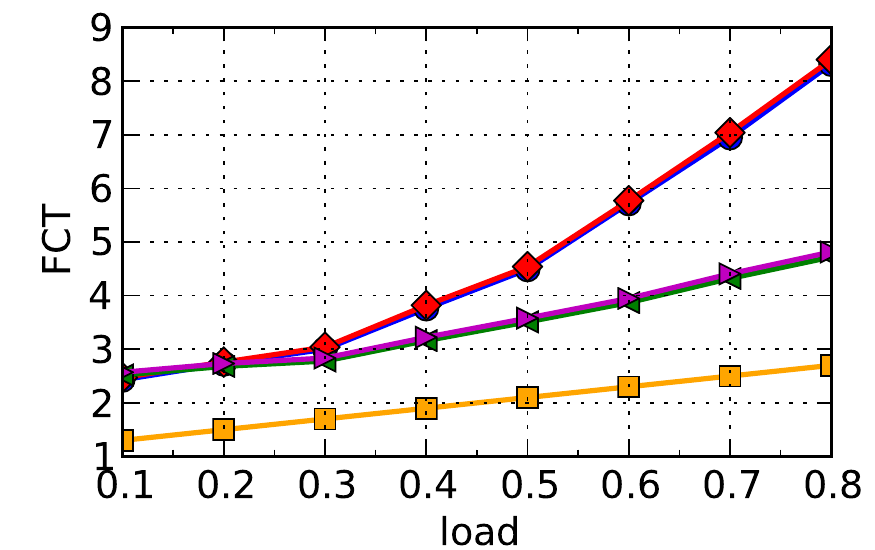}
	\vspace{-4.5mm}
	\caption{(100KB, $\infty$): Avg}
	\label{fig:mean-large-web-sim}
\end{subfigure}
\caption{FCT breakdown for different flows with a 16-pod Fat-tree and the web search workload \cite{AGMP10} in NS-3. } \label{fig:web-sim}
\vspace{-3mm}
\end{figure*}
\begin{figure*}[t]
\begin{subfigure}[h]{0.33\linewidth}
	\centering
	\includegraphics[width=1\linewidth]{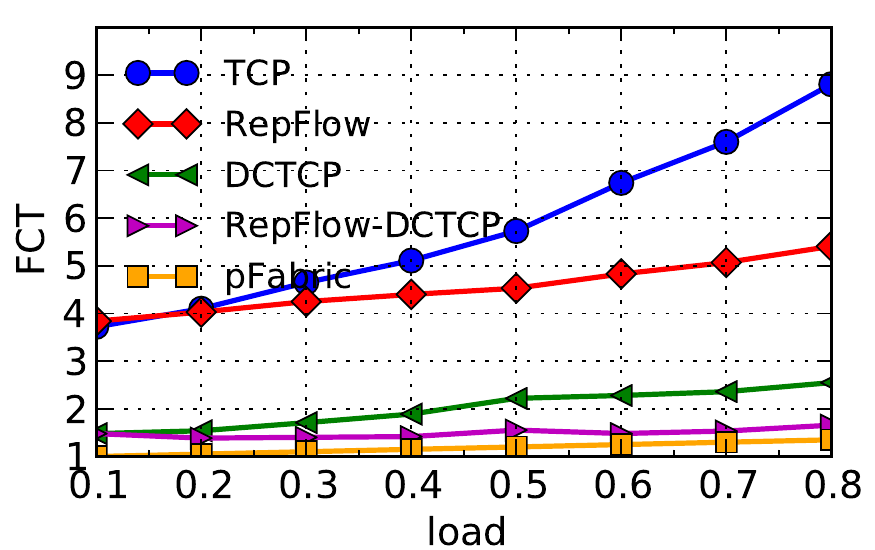}
	\vspace{-4.5mm}
	\caption{(0,100KB]: Avg}
	\label{fig:mean-short-db-sim}
\end{subfigure}
\begin{subfigure}[h]{0.33\linewidth}
	\centering
	\includegraphics[width=1\linewidth]{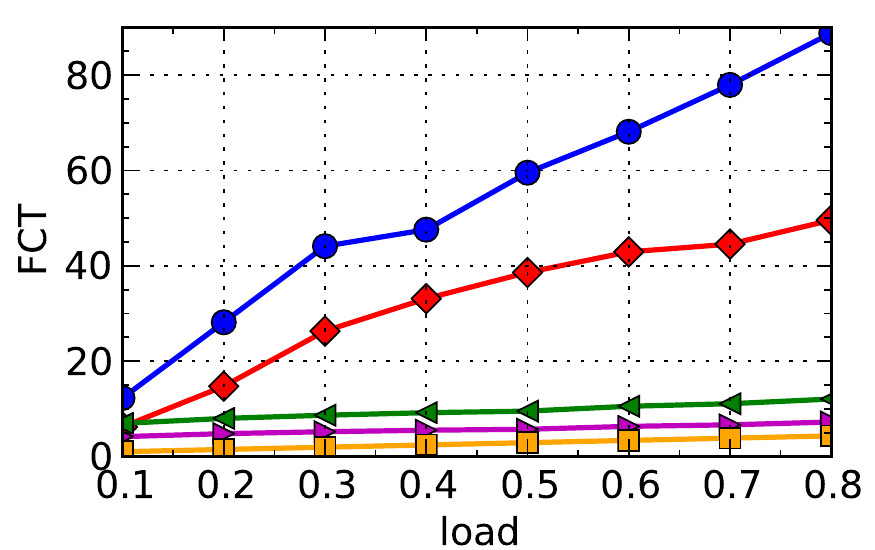}
	\vspace{-4.5mm}
	\caption{(0,100KB]: 99-th percentile}
	\label{fig:tail-short-db-sim}
\end{subfigure}
\begin{subfigure}[h]{0.33\linewidth}
	\centering
	\includegraphics[width=1\linewidth]{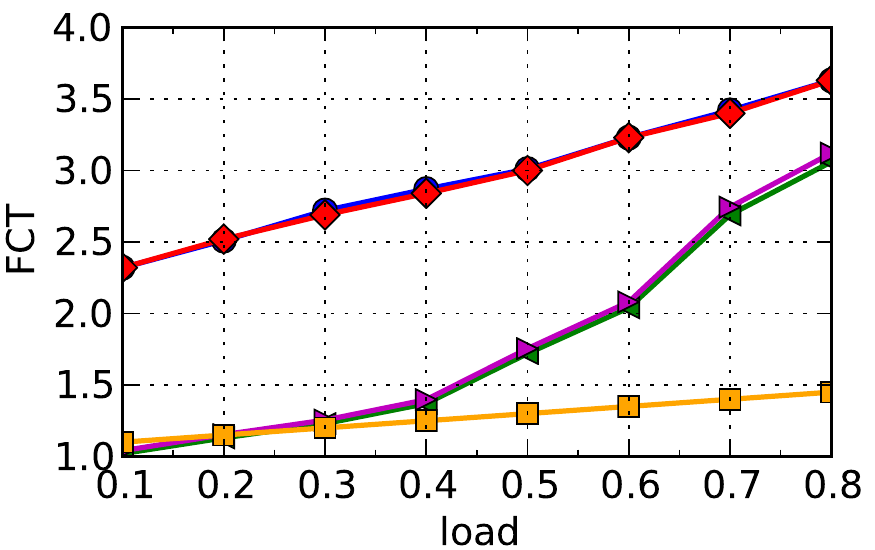}
	\vspace{-4.5mm}
	\caption{(100KB, $\infty$): Avg}
	\label{fig:mean-large-db-sim}
\end{subfigure}
\caption{FCT breakdown for different flows with a 16-pod Fat-tree and the data mining workload \cite{GHJK09} in NS-3. } \label{fig:db-sim}
\vspace{-3mm}
\end{figure*}

\subsection{Summary}
We summarize our analytical findings. Short flows' mean and tail FCT depend critically on queueing delay, and the factor $\frac{\rho}{1-\rho}$ assuming a M/G/1-FCFS queue. Using replication, they have much less probability of entering a busy queue, and the effective load they experience is greatly reduced. This confirms the intuition that RepFlow provides path diversity gains in data center networks. RepFlow is expected to have speedup around 40\%--70\% as numerical results show. The negative impact on large flow is very mild, because from large flows' perspectives, load only increases slightly.

\section{Experimental Evaluation}
\label{sec:evaluation}

We now evaluate RepFlow's performance using extensive packet-level simulations in the NS-3 simulator. Building on this, we show how RepFlow performs in a realistic small-scale implementation running Linux kernel code based on Mininet \cite{HHJL12} in the next section.

\subsection{Methodology}
\label{sec:sim_setup}

{\bf Topology:} We use a 16-pod Fat-tree as the network topology \cite{ALV08}, which is commonly used in data centers. The fabric consists of 16 pods, each containing an edge layer and an aggregation layer with 8 switches each. The edge switches in each pod connect to 8 hosts each. The network has 1,024 hosts and 64 core switches. There are 64 equal-cost paths between any pair of hosts at different pods. Each switch is a 16-port 1Gbps switch, resulting in a full bisection bandwidth network. The end-to-end round-trip latency is $\sim$32$\mu$s. ECMP is used as the load balancing scheme. 

{\bf Benchmark workloads:} We use empirical workloads to reflect traffic patterns that have been observed in production data centers. We consider two flow size distributions. The first is from a cluster running web search \cite{AGMP10}, and the second is from a data center mostly running data mining jobs \cite{GHJK09}. Both workloads exhibit heavy-tailed characteristics with a mix of small and large flows. In the web search workload, over 95\% of the bytes are from 30\% of flows larger than 1MB. In the data mining workload, 95\% of all bytes are from $\sim$3.6\% flows that are larger than 35MB, while more than 80\% of flows are less than 10KB. Flows are generated between random pairs of hosts following a Poisson process with load varying from 0.1 to 0.8 to thoroughly evaluate RepFlow's performance in different traffic conditions. We simulate 0.5s worth of traffic for the network at each run, and ten runs for each load. The entire simulation takes around 900+ machine-hours. 

\subsection{Schemes Compared}

{\bf TCP:} Standard TCP-New Reno is used as the baseline of our evaluation. The initial window is set to 12KB, and switches use DropTail queues with a buffer size of 100 packets. These are standard settings used in many studies \cite{VHV12,AYSK13}.

{\bf RepFlow:} Our design as described in Sec.~\ref{sec:design}. All flows less than 100KB are replicated. Other parameters are the same as TCP.

{\bf DCTCP:} The DCTCP protocol with ECN marking at DropTail queues \cite{AGMP10}. Our implementation is based on a copy of the source code we obtained from the authors of D2TCP \cite{VHV12}. The ECN marking threshold is set to 5\%. Other parameters are set following \cite{AGMP10}.

{\bf RepFlow-DCTCP:} This is RepFlow on top of DCTCP. As discussed in Sec.~\ref{sec:design} RepFlow can work with any TCP variant. We use this scheme to demonstrate RepFlow's ability to further reduce FCT for networks that have already modified hosts to adopt specialized data center transport such as DCTCP.

{\bf pFabric:} This is the state-of-the-art approach that offers near-optimal performance in terms of minimizing flow completion times \cite{AYSK13}. pFabric assigns higher priority to flows with less remaining bytes to transfer, encodes the flow priority in packet headers, and modifies switches so they schedule packets based on flow priority. Thus short flows are prioritized with near-optimal FCT. Our implementation is based on a copy of the source code we obtained from the authors of the paper \cite{AYSK13}. We follow \cite{AYSK13} and set the DropTail queue size to be 36KB at each switch port for best performance. 


\subsection{RepFlow on TCP}
\label{sec:ns3-fct}

We first evaluate RepFlow's performance with TCP. RepFlow significantly reduces the mean and 99-th percentile FCT for short flows in both workloads compared to TCP. Fig.~\ref{fig:web-sim} and Fig.~\ref{fig:db-sim} show the FCT performance for different flows in the web search and data mining workloads as we vary the load from 0.1 to 0.8. With the web search workload, RepFlow reduces mean FCT by $\sim$40\%--45\% for short flows for loads from 0.4 to 0.8. The improvement in tail FCT is more salient. RepFlow can be $\sim$10x faster than TCP when the load is low ($<$0.4), and over 60\% in all other loads. The data mining workload yields qualitatively similar observations. 

The results demonstrate the advantage of replication in harvesting the multi-path diversity, especially when the load is relatively low which is usually the case in production networks \cite{KSGP09}. The tail FCT reduction is more substantial in low loads, while the mean FCT reduction is more significant in high loads. The reason is that when the load is low, most short flows finish adequately, with a few exceptions that hit the long tail. Since path diversity is abundant with low loads, RepFlow can easily reduce the tail latency by a large margin. When the load is higher, almost every short flow experiences some queueing delay resulting in longer FCT in the average case. RepFlow is thus able to provide diversity gains even in the average case for most short flows. The results also corroborate our analysis and numerical results in Sec.~\ref{sec:model} despite numerical differences.

We also look at the impact of replication on large flows. From Fig.~\ref{fig:mean-large-web-sim} and \ref{fig:mean-large-db-sim} we can see that large flows suffer negligible FCT increment. Thus in a realistic network with production workloads, RepFlow causes little performance degradation and perform even better than our analysis predicts. Note that large flows in the data mining workload has better FCT than those in the web search workload. This is because the data mining workload is more skewed with elephant flows larger than 1MB, while in the web search workload there are many ``medium'' flows of size between 100KB and 1MB. These medium flows also suffer from queueing delay especially in the slow-start phase, resulting in the larger FCT in Fig.~\ref{fig:mean-large-web-sim}.

\subsection{RepFlow on DCTCP}

\begin{figure}[ht]
\begin{subfigure}[h]{0.49\linewidth}
	\centering
	\includegraphics[width=1\linewidth]{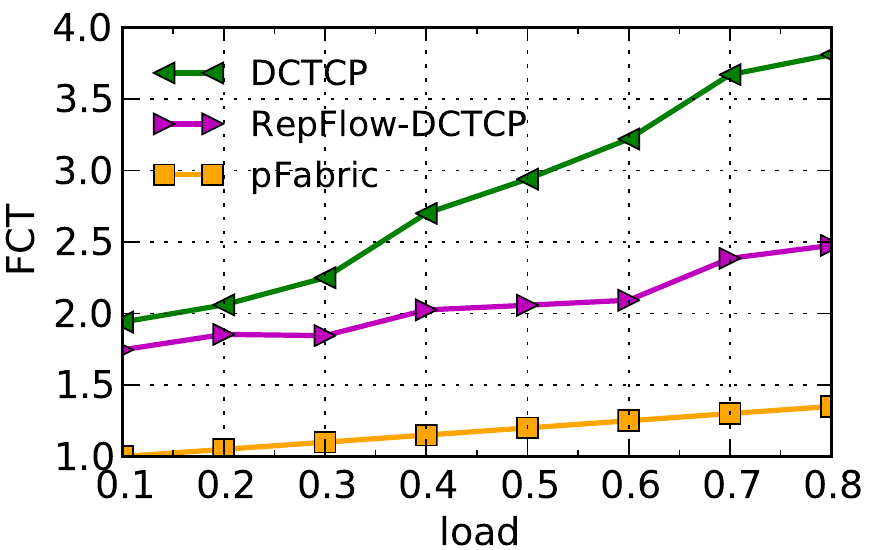}
	\vspace{-4.5mm}
	\caption{Web search workload.}
	\label{fig:mean-web-dctcp}
\end{subfigure}
\begin{subfigure}[h]{0.49\linewidth}
	\centering
	\includegraphics[width=1\linewidth]{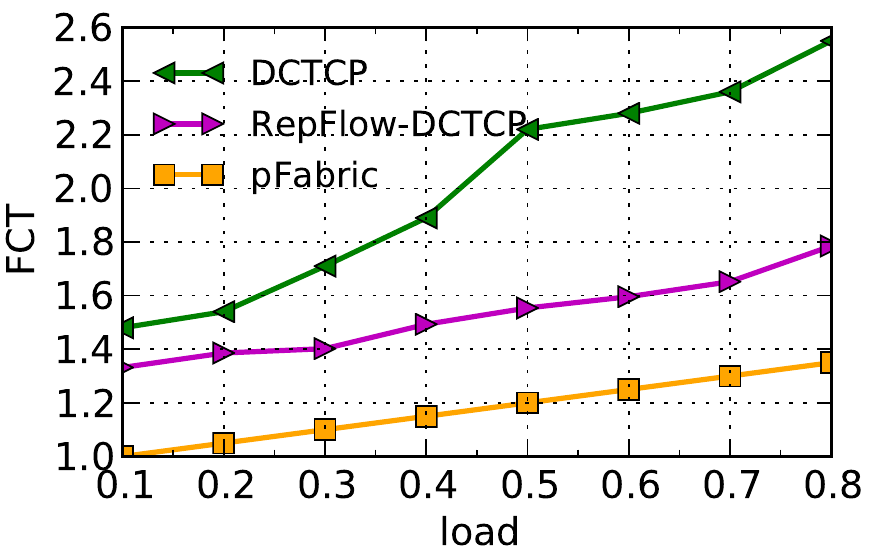}
	\vspace{-4.5mm}
	\caption{Data mining workload.}
	\label{fig:mean-db-dctcp}
\end{subfigure}
\caption{Mean FCT for short flows with a 16-pod Fat-tree in NS-3. Note the different ranges of the y-axis in the plots.}
\label{fig:mean-zoomin}
\end{figure}

\begin{figure}[ht]
\begin{subfigure}[h]{0.49\linewidth}
	\centering
	\includegraphics[width=1\linewidth]{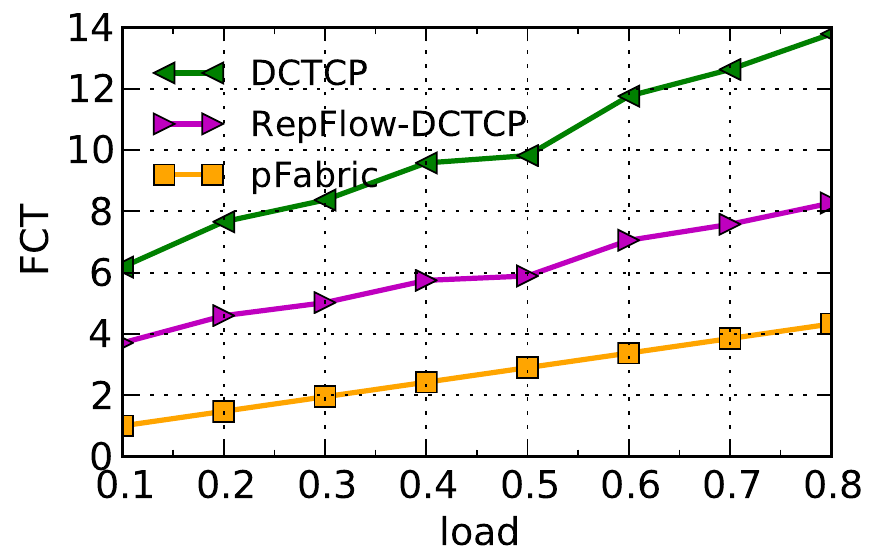}
	\vspace{-4.5mm}
	\caption{Web search workload.}
	\label{fig:tail-web-dctcp}
\end{subfigure}
\begin{subfigure}[h]{0.49\linewidth}
	\centering
	\includegraphics[width=1\linewidth]{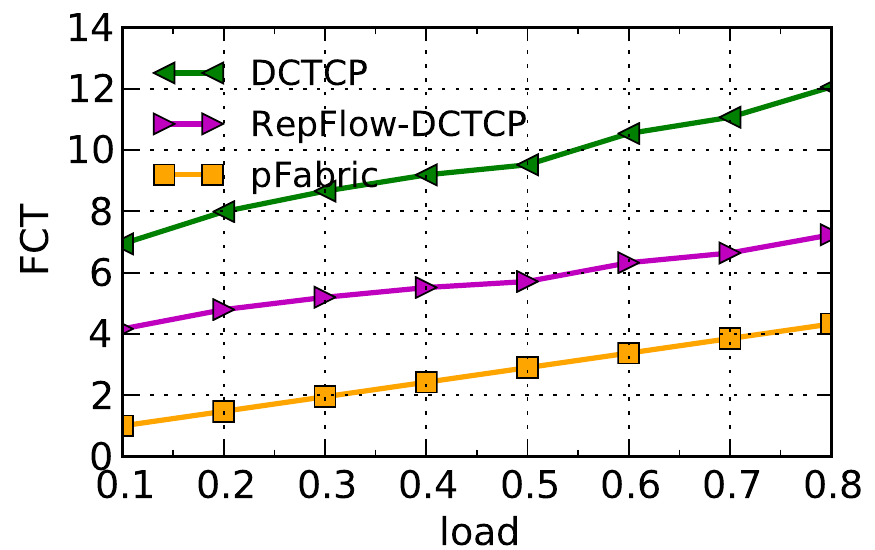}
	\vspace{-4.5mm}
	\caption{Data mining workload.}
	\label{fig:tail-db-dctcp}
\end{subfigure}
\caption{99-th percentile FCT for short flows with a 16-pod Fat-tree.}
\label{fig:tail-zoomin}
\end{figure}

RepFlow's full potential is realized when used together with specialized data center transport such as DCTCP \cite{AGMP10}. For data centers that have already adopted DCTCP, RepFlow can be readily utilized just as with regular TCP. We observe in Fig.~\ref{fig:web-sim} and Fig.~\ref{fig:db-sim} that DCTCP improves FCT significantly compared to TCP, especially the 99-th percentile FCT for short flows. We show the mean and tail FCT results with only DCTCP, RepFlow-DCTCP and pFabric in Fig.~\ref{fig:mean-zoomin} and \ref{fig:tail-zoomin} for better contrasts. Observe that RepFlow-DCTCP reduces mean FCT by another 40\% in both workloads, and is only $\sim$30\% slower than pFabric for the data mining workload. In terms of 99-th percentile FCT, DCTCP, RepFlow-DCTCP and pFabric all provide almost an order of magnitude reduction compared to TCP. RepFlow cuts down another $\sim$35\% tail FCT on top of DCTCP, and performs close to the near-optimal pFabric \cite{AYSK13}. In general short flows perform better in the data mining workload than in the web search workload. This is because in the data mining workload it is less likely that multiple large flows are transmitting concurrently on the same port, implying less contention with the short flows.

The improvements of RepFlow with DCTCP are less significant than with regular TCP. The reason is that by using ECN marking, DCTCP keeps the queue length very small most of the time, yielding less path diversity for RepFlow to exploit. 

\vspace{-3mm}

\begin{figure}[hpt]
	\centering
\begin{subfigure}[h]{0.49\linewidth}
	\centering
	\includegraphics[width=1\linewidth]{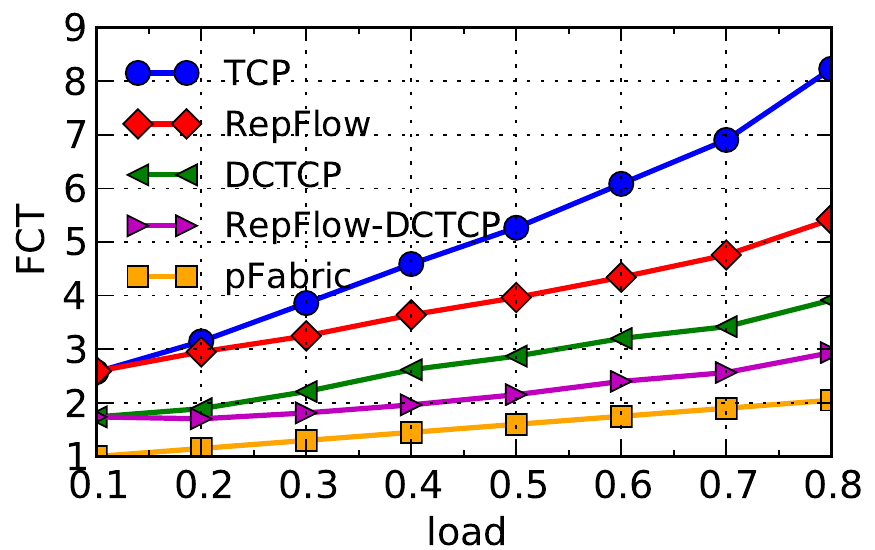}
	\vspace{-4.5mm}
	\caption{Web search workload}
	\label{fig:mean-total-web-sim}
\end{subfigure}
\begin{subfigure}[h]{0.49\linewidth}
	\centering
	\includegraphics[width=1\linewidth]{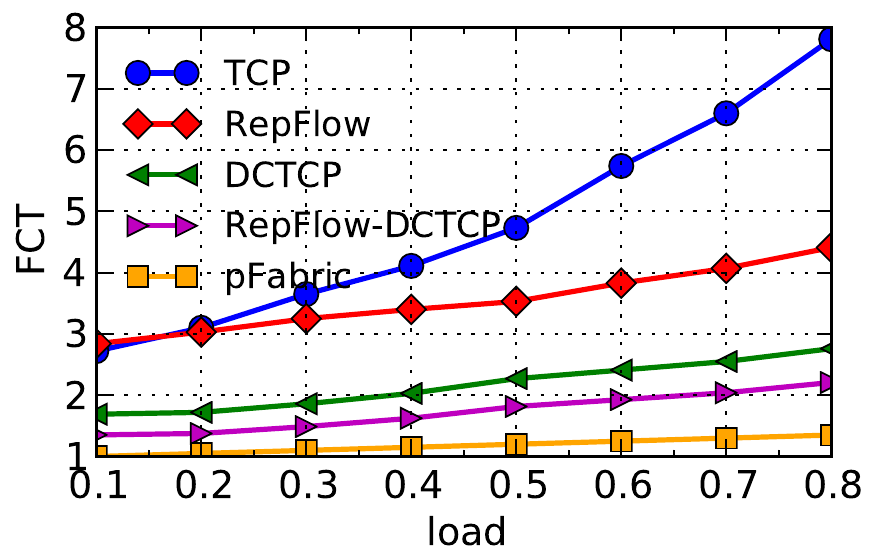}
	\vspace{-4.5mm}
	\caption{Data mining workload}
	\label{fig:mean-total-db-sim}
\end{subfigure}
\caption{Mean FCT for all flows.}
\vspace{-4mm}
\label{fig:overall}
\end{figure}

Overall, Fig.~\ref{fig:overall} shows the average FCT across all flows for all schemes. RepFlow improves TCP by $\sim$30\%--50\% in most cases. RepFlow-DCTCP improves DCTCP further by $\sim$30\%, providing very close-to-optimal FCT compared to state-of-the-art pFabric.

\subsection{Replication Overhead}
\label{sec:overhead}
Replication clearly adds more traffic to the network. In this section we investigate the overhead issue of RepFlow. We calculate the percentages of extra bytes caused by replicated flows in both workloads for all loads as shown in Table~\ref{table:overhead}. Since short flows only generate a tiny fraction of all bytes in both workloads, not surprisingly replication does not incur excessive overhead to the network. The overhead for the DCTCP implementation is essentially the same and we omit the results. 

\begin{table}[htp]
\begin{subtable}[ht]{1\linewidth}
    \centering
    \begin{tabular}{  c |c|c|c|c|c|c|c }
    \hline 
         0.1 & 0.2 & 0.3 & 0.4 & 0.5 & 0.6 & 0.7 & 0.8  \\ \hline
         3.45\% & 2.78\% & 3.13\% & 3.38\% & 3.29\% & 3.47\% & 3.22\% & 3.27\% \\ \hline
    \end{tabular}
    \caption{Web search workload}
    \label{table:overhead-web}
    \end{subtable}
    ~
    \begin{subtable}[ht]{1\linewidth}
    \centering
    \begin{tabular}{  c |c|c|c|c|c|c|c }
    \hline 
         0.1 & 0.2 & 0.3 & 0.4 & 0.5 & 0.6 & 0.7 & 0.8  \\ \hline
         1.41\% & 1.18\% & 2.13\% & 1.38\% & 1.33\% & 1.07\% & 1.12\% & 1.09\% \\ \hline
    \end{tabular}
    \caption{Data mining workload}
    \label{table:overhead-db}
    \end{subtable}
\caption{Overhead of RepFlow in NS-3 simulation.}
\label{table:overhead}
\end{table}

To summarize, RepFlow achieves much better FCT for short flows compared to TCP with minimum impact on large flows. The improvements are not as significant as pFabric \cite{AYSK13}. This is expected since RepFlow only opportunistically utilizes the less congested path without being able to reduce switch queueing delay. However since RepFlow does not require switch hardware or kernel changes, it represents an effective and practical approach to the imminent problem of reducing FCT. On the other hand when it is feasible to implement RepFlow on top of advanced protocols such as DCTCP, RepFlow performs competitively compared to pFabric, again without the burden of modifying switch hardware.


\section{Implementation on Mininet}
\label{sec:mininet}

We implement RepFlow on Mininet, a high-fidelity network emulation framework based on Linux container based virtualization \cite{HHJL12}. Mininet creates a realistic virtual network, running real Linux kernel, switch and application code on a single machine. Though its scale is smaller than production data center networks due to the single-machine CPU limitation (tens of Mbps link bandwidth compared to 1Gbps), it has been shown to faithfully reproduce implementation results from \cite{AGMP10,ARRH10} with high fidelity \cite{HHJL12}, and has been used as a flexible testbed for networking experiments \cite{KZZC13}.

Our implementation follows the design in Sec.~\ref{sec:design}. We run socket-based sender and receiver programs as applications on virtual hosts in Mininet. Each virtual host runs two receivers, one for receiving regular flows and the other for replicated flows, in separate threads and listening on different ports. A flow is created by spawning a sender thread that sends to the regular port of the receiving host, and if it is a short flow another sender thread sending to the other port. The replicated flow shares the same source port as the original flow for easy idenfitification. We implement RepFlow on top of TCP-New Reno in Mininet 2.0.0 on a Ubuntu 12.10 LTS box. 

We use a 4-pod Fat-tree topology with 16 hosts connected by 20 switches, each with 4 ports. Each port has 50 packets of buffer. We set link bandwidth to 20Mb and delay to 1ms. The minimum delay Mininet supports is 1ms without high-precision timers. RiplPOX is installed as the controller on switches to support ECMP. The entire experiment takes $\sim$6 hours to run on an EC2 c1.xlarge instance with 8 virtual cores. We observe that Mininet becomes unstable when the load exceeds 0.5, possibly due to its scalability limitation. Thus we only show results for loads from 0.1 to 0.5.

\begin{figure*}[ht]
\begin{subfigure}[h]{0.24\linewidth}
	\centering
	\includegraphics[width=1\linewidth]{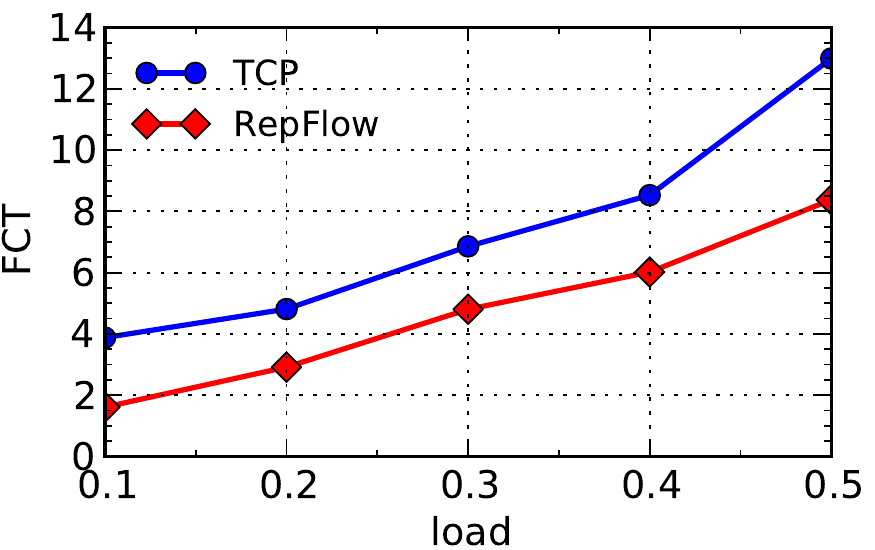}
	\vspace{-4.5mm}
	\caption{(0, 100KB]: Avg}
	\label{fig:mean-short-web}
\end{subfigure}
\begin{subfigure}[h]{0.24\linewidth}
	\centering
	\includegraphics[width=1\linewidth]{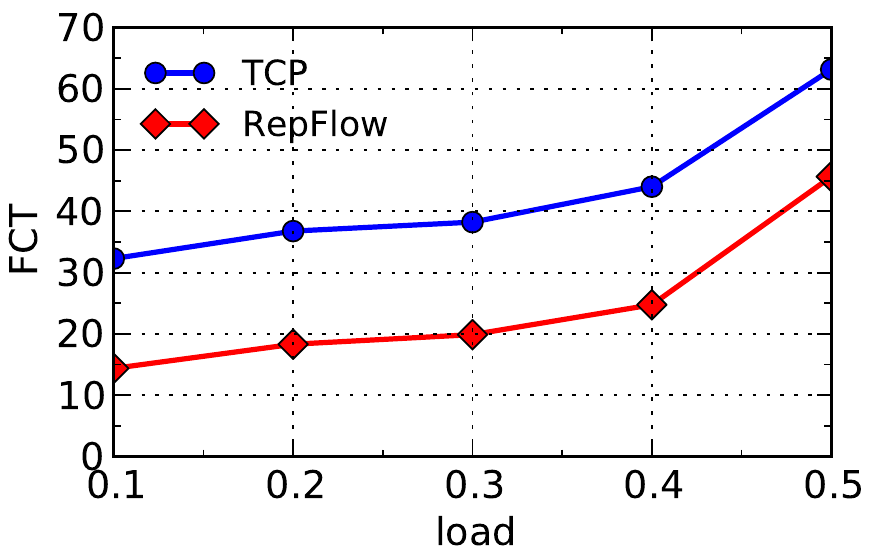}
	\vspace{-4.5mm}
	\caption{(0, 100KB]: 99-th percentile}
	\label{fig:tail-short-web}
\end{subfigure}
\begin{subfigure}[h]{0.24\linewidth}
	\centering
	\includegraphics[width=1\linewidth]{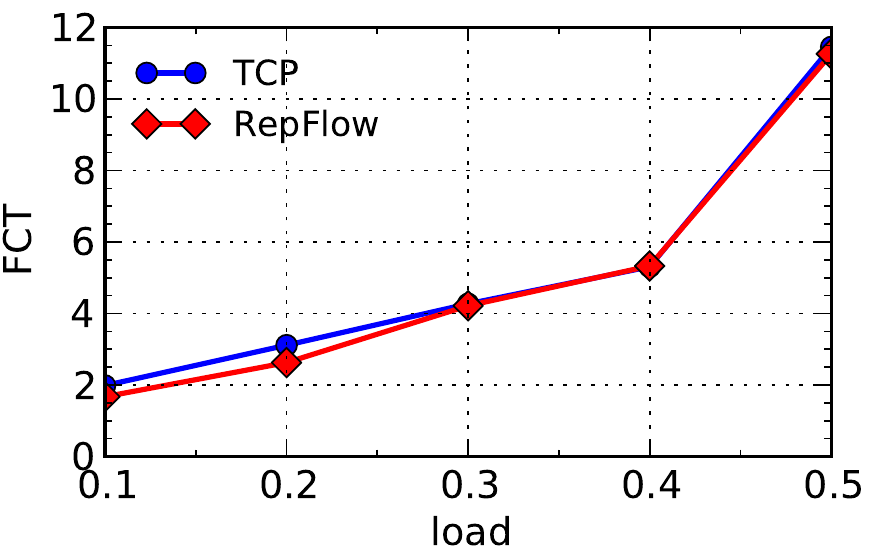}
	\vspace{-4.5mm}
	\caption{(100KB, $\infty$): Avg}
	\label{fig:mean-large-web}
\end{subfigure}
\begin{subfigure}[h]{0.24\linewidth}
	\centering
	\includegraphics[width=1\linewidth]{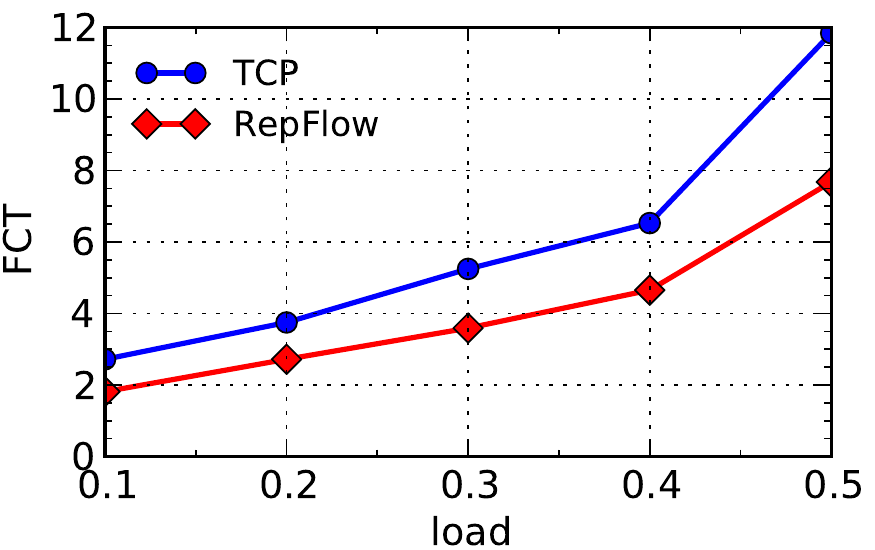}
	\vspace{-4.5mm}
	\caption{Overall: Avg}
	\label{fig:mean-total-web}
\end{subfigure}
\caption{Implementation on Mininet with a 4-pod Fat-tree and the web search workload \cite{AGMP10}.}\label{fig:mininet-web}
\vspace{-3mm}
\end{figure*}

\begin{figure*}[ht]
\begin{subfigure}[h]{0.24\linewidth}
	\centering
	\includegraphics[width=1\linewidth]{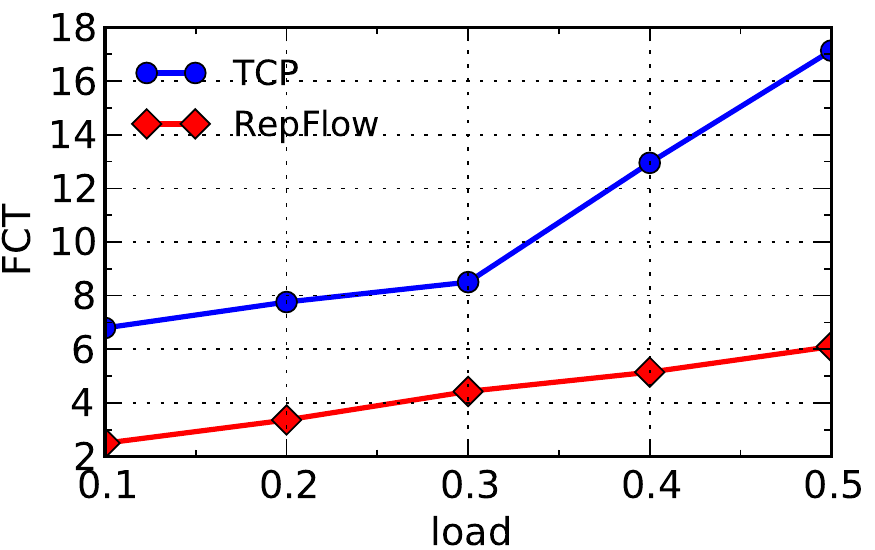}
	\vspace{-4.5mm}
	\caption{(0, 100KB]: Avg}
	\label{fig:mean-short-db}
\end{subfigure}
\begin{subfigure}[h]{0.24\linewidth}
	\centering
	\includegraphics[width=1\linewidth]{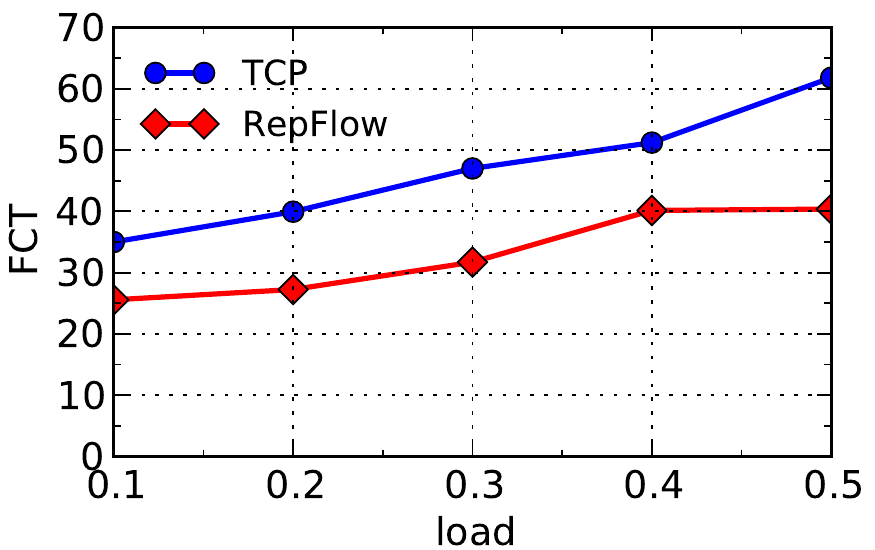}
	\vspace{-4.5mm}
	\caption{(0, 100KB]: 99-th percentile}
	\label{fig:tail-short-db}
\end{subfigure}
\begin{subfigure}[h]{0.24\linewidth}
	\centering
	\includegraphics[width=1\linewidth]{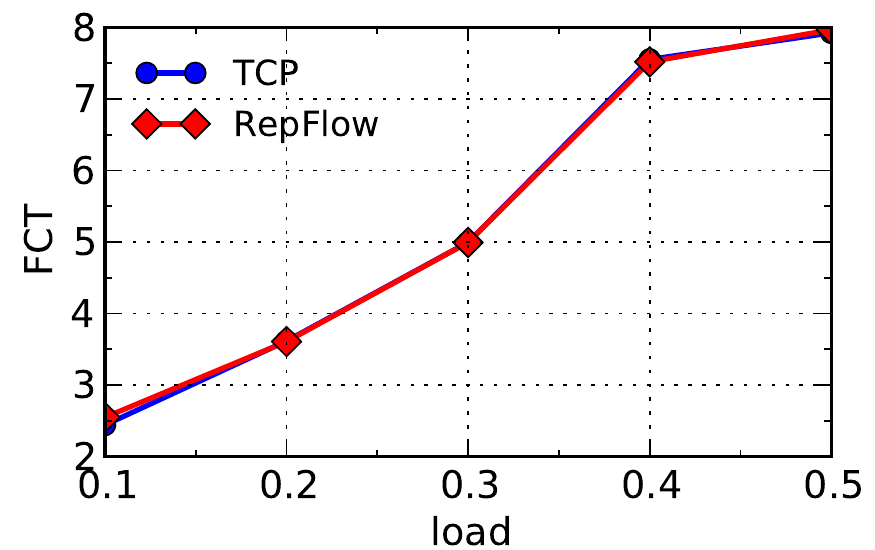}
	\vspace{-4.5mm}
	\caption{(100KB, $\infty$): Avg}
	\label{fig:mean-large-db}
\end{subfigure}
\begin{subfigure}[h]{0.24\linewidth}
	\centering
	\includegraphics[width=1\linewidth]{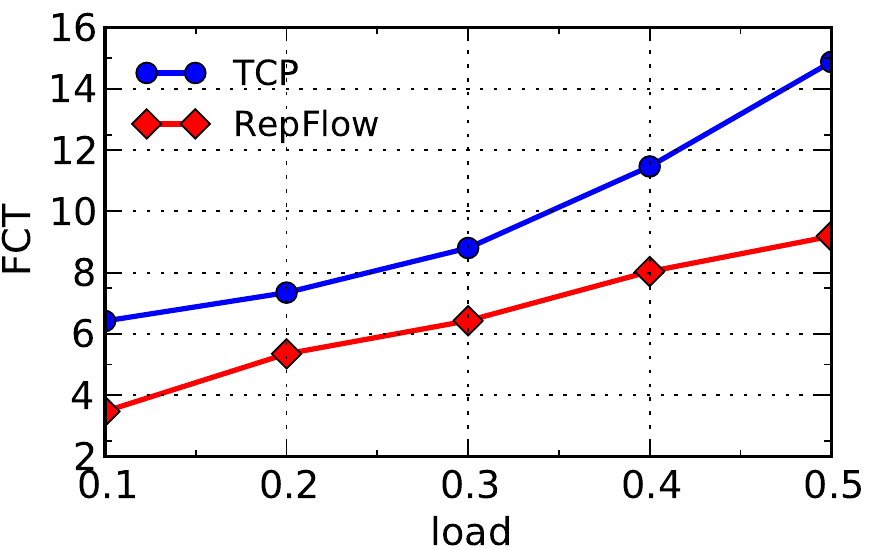}
	\vspace{-4.5mm}
	\caption{Overall: Avg}
	\label{fig:mean-total-db}
\end{subfigure}
\caption{Implementation on Mininet with a 4-pod Fat-tree and the data mining workload \cite{GHJK09}.}\label{fig:mininet-db}
\vspace{-3mm}
\end{figure*}

Fig.~\ref{fig:mininet-web} and \ref{fig:mininet-db} show the results. Fig.~\ref{fig:mean-short-web} and \ref{fig:mean-short-db} show RepFlow has $\sim$25\%--50\% and 50\%--70\% mean FCT improvement in the web search and data mining workload, respectively. The improvement in tail FCT is around 30\% in most cases for both workloads and is smaller than the NS-3 simulation results. The reason is two-fold. First there are fewer equal-cost paths in the 4-pod Fat-tree than the 16-pod Fat-tree in the simulation, implying less path diversity for RepFlow. Second, in the Mininet implementation, each sender thread of the replicated flow is forked after the original flow. The time difference due to virtualization overhead is non-negligible for short flows ($\sim$1ms). Thus it less likely for the replicated flows to finish faster, reducing the potential of RepFlow. We believe in a real implementation without virtualization this is unlikely to become an issue. Fig.~\ref{fig:mean-large-web} and \ref{fig:mean-large-db} confirm that large flows are not affected by replication. Overall the implementation results are in line with simulation results. 

\section{Concluding Remarks}

We presented the design and evaluation of RepFlow, a simple approach that replicates short TCP flows to reap path diversity in data centers to minimize flow completion times. Analytical and experimental results demonstrate that it reduces mean and tail FCT significantly with no changes to existing infrastructures. We believe flow replication is an intuitive approach to combat unpredictable performance degradations, including but not limited to slow and long-tailed FCT. 

Our work is a first step in this direction. Our next step is to prototype RepFlow as a general application library running on TCP and DCTCP, and evaluate its benefits for real applications in a deployed data center. Ultimately, this may pave the path for practical use of RepFlow at scale. Many interesting issues remain. For example, intuitively RepFlow improves resilience against failures which needs to be better understood. A more detailed theoretical model of TCP flow completion times in data center networks would also provide a principled analysis of RepFlow's benefits.

\section*{Acknowledgment}
We thank Balajee Vamanan and Shuang Yang for providing their simulation code in the papers \cite{VHV12} and \cite{AYSK13} for DCTCP and pFabric, respectively. We also thank David Maltz and Christopher Stewart for providing positive feedback and suggestions in the early stage of this work.
\bibliographystyle{abbrv}
\bibliography{IEEEabrv,main}

\end{document}